\documentclass[aps,twocolumn,pra,superscriptaddress,showpacs,letterpaper,nofootinbib]{revtex4}
\usepackage{amsmath,amssymb}
\usepackage{color}
\usepackage[pdftex]{graphicx}
\usepackage{hyperref}
\usepackage{mathptmx}

\begin{document}

\newcommand{\be}{\begin{equation}}
\newcommand{\ee}{\end{equation}}
\newcommand{\R}[1]{\textcolor{red}{#1}}
\long\def\symbolfootnote[#1]#2{\begingroup%
\def\thefootnote{\fnsymbol{footnote}}\footnote[#1]{#2}\endgroup}


\title{Quantum back-action in measurements of zero-point mechanical oscillations}

\author{Farid Ya.\ Khalili}
\affiliation{Physics Faculty, Moscow State University, Moscow
119991, Russia}

\author{Haixing Miao}
\affiliation{Theoretical Astrophysics 350-17, California Institute
of Technology, Pasadena, CA 91125, USA}

\author{Huan Yang}
\affiliation{Theoretical Astrophysics 350-17, California Institute
of Technology, Pasadena, CA 91125, USA}

\author{Amir H. Safavi-Naeini}
\affiliation{Thomas J. Watson, Sr., Laboratory of Applied Physics,
California Institute of Technology, Pasadena, California 91125, USA}

\author{Oskar Painter}
\affiliation{Thomas J. Watson, Sr., Laboratory of Applied Physics,
California Institute of Technology, Pasadena, California 91125, USA}

\author{Yanbei Chen}
\affiliation{Theoretical Astrophysics 350-17, California Institute
of Technology, Pasadena, CA 91125, USA}


\begin{abstract}
Measurement-induced back action, a direct consequence of the Heisenberg
Uncertainty Principle, is the defining feature of quantum measurements.
We use quantum measurement theory to analyze the recent experiment of
Safavi-Naeini {\it et al.} [Phys. Rev. Lett. {\bf 108}, 033602 (2012)], and
show that results of this experiment not only characterize the zero-point
fluctuation of a near-ground-state nanomechanical oscillator, but also
demonstrate the existence of quantum back-action noise --- through
correlations that exist between sensing noise and back-action noise. These
correlations arise from the quantum coherence between the mechanical
oscillator and the measuring device, which build up during the measurement
process, and are key to improving sensitivities beyond the Standard Quantum
Limit.
\end{abstract}

\maketitle


\section{Introduction}
\label{introduction}


Quantum mechanics dictates that no matter or field can stay absolutely at rest,
even at the ground state, for which energy is at minimum.  A starting point for
deducing this inevitable fluctuation is to write down the Heisenberg Uncertainty
Principle
\begin{equation}
\left[\hat x, \,\hat p\right]=i\,\hbar \,,
\end{equation}
which leads to:
\begin{equation}
\label{hei1}
\Delta x \cdot \Delta p \ge \hbar/2\,.
\end{equation}
Here $\hat x$ and $\hat p$ are the position and momentum operators, while
$\Delta x$ and $\Delta p$ are standard deviations of position and momentum
for an arbitrary quantum state.  Eq.\,\eqref{hei1} means we cannot specify
the position and momentum of a harmonic oscillator simultaneously, as a point
in classical phase space --- the oscillator must at least occupy $\hbar/2$ area in
the phase space. If the oscillator has mass of $m$ and eigenfrequency of
$\omega_m$, then in the Heisenberg picture we can write
\begin{equation}
\label{comm}
\left[\hat x_q(t),\,\hat x_q(t')\right] = \frac{i \hbar\sin\omega_m(t'-t) }{m\omega_m}\,,
\end{equation}
which leads to:
\begin{equation}
\label{heis2}
\Delta x_q(t) \cdot \Delta x_q(t') \ge \frac{\hbar|\sin\omega_m(t'-t)| }{2m\omega_m}\,.
\end{equation}
with $\hat x_q(t)$ being the Heisenberg operator of the oscillator position,
quantum-mechanically evolving under the free Hamiltonian.
Here $\Delta x_q(t)$ is the standard deviation of $\hat x_q(t)$
for an arbitrary quantum state.
Eq.\,\eqref{heis2} means the position of a freely
evolving quantum harmonic oscillator cannot continuously assume precise values,
but instead, must fluctuate. This fluctuation carries the zero-point mechanical
energy of $\hbar\omega_m/2$.

As a key feature of quantum mechanics, zero-point fluctuation of displacement
is an important effect to verify when we bring macroscopic mechanical degrees
of freedom into their ground states~\cite{Kippenberg2008, Marquardt2009,
Connell2010,Verlot2009,Teufel2009, Anetsberger2009, Borkje2010, Westphal2011}.
Needless to say, a continuous observation of the zero-point fluctuation
of a macroscopic mechanical oscillator requires superb displacement sensitivity.

However,  what constitutes an ``observation of the quantum zero-point fluctuation''
is conceptually subtle.  Eqs.\,\eqref{comm} and \eqref{heis2}, which argue
for the inevitability of the zero-point fluctuation, also dictate that the ``exact amount''
of the zero-point fluctuation cannot be determined precisely.  More specifically, if we
use a linear measurement device to probe the zero-point fluctuation, which has an output
field of  $\hat y(t)$, then we must at least have
\begin{equation}
\label{vanishcomm}
\left[\hat y(t),\,\hat y(t')\right]=0
\end{equation}
at all times, in order for $\hat y(t)$ to be able to represent experimental data string--- with measurement
noise simply due to the projection of the device's quantum state into simultaneous eigenstates
of all $\{\hat y(t):\;t\in \mathbb{R}\}$.  This means $\hat y$ must be written as
 \begin{equation}
 \label{eqy1}
 \hat y(t) = \hat \epsilon(t)  + \hat x_{q}(t)
 \end{equation}
with non-vanishing additional noise (error) $\hat \epsilon(t)$, which consists of degrees
of freedom of the measurement device and compensates the non-vanishing commutator
of $\hat x_{q}$\,\symbolfootnote[3]{We note that Ozawa has developed a different formalism
to quantify the issues that arise when attempts are made to measure non-commuting
observables like $\hat x_q(t)$\,\cite{Ozawa2002,Sponar:2012vv}. However, we
have chosen to adopt the Braginsky-Khalili approach\,\cite{BK92}, because it is immediately
applicable when the non-commuting observable is acting as a probe for an external classical force.}.
In addition, during the measurement process, actual evolution of
the mechanical displacement $\hat x$ must differ from its free evolution $\hat x_q$.
This is because
\begin{equation}
\frac{\left[\hat x(t),\hat x(t')\right]}{i\hbar } \equiv \chi(t'-t)
\end{equation}
is also the classical response function of $x$ to an external force: any device
that attempts to measure $\hat x$ by coupling it with an external observable
$\hat F$, which introduces a term proportional to $\hat x\, \hat F$ into the Hamiltonian, 
will have to cause non-zero disturbance.  For this reason, we can expand the measurement
error $\hat \epsilon$ into two parts: $\hat z$ --- the sensing noise that is independent from
mechanical motion and $\hat x_{\rm BA}$ --- additional disturbance to the mechanical
motion from the measurement-induced back-action, and rewrite $\hat y(t)$ as:
 \begin{equation}
 \label{eqy2}
 \hat y(t) =\underbrace{\hat z(t)  + \hat x_{\rm BA} (t)}_{\hat \epsilon(t)}+ \hat x_q(t)=\hat z(t)+\hat x(t)\,.
 \end{equation}
The mechanical displacement under measurement is therefore a sum of the freely-evolving
operator $\hat x_q$ plus the disturbance $\hat x_{\rm BA}$ due to back action noise, namely,
$\hat x(t)=\hat x_q(t)+\hat x_{\rm BA}(t)$.

The above lines of reasoning lie very much at the heart of linear quantum measurement
theory, pioneered by Braginsky in the late 1960s aiming at describing resonant-bar
gravitational-wave detectors\,\cite{SQL1,BK92}, and later adapted to the analysis of laser
interferometer gravitational-wave detectors by Caves~\cite{Caves1980}.  A key concept
in linear quantum measurement theory is the trade-off between sensing noise and
back-action noise, which gives rise to the so-called Standard Quantum Limit (SQL).
For optomechanical devices, sensing noise takes the form of quantum shot noise due to
discreteness of photons, while the quantum back-action is enforced by quantum fluctuations
in the radiation pressure acting on the mechanical oscillators~\cite{Caves1980}, which is
therefore also called quantum radiation-pressure noise. It has been shown that the SQL,
although not a strict limit for sensitivity, can only be surpassed by carefully designed
linear measurement devices which take advantage of quantum correlations between
the sensing noise and the back-action noise.

Observing signatures of quantum back-action, achieving and surpassing the associated
SQL in mechanical systems are of great importance for the future of quantum-limited
metrology, e.g., gravitational-wave detections~\cite{klmtv,Chen1,Purdue2,Corbitt2006,
Corbitt2007,Corbitt2007b,Marino2010, ChenDanilishin2011,Danilishin2012}. At the
moment, it is still experimentally challenging to directly observe quantum radiation-pressure
noise in optomechanical devices due to high levels of environmental thermal fluctuations, and
there are significant efforts being made toward this\,\cite{Verlot2009, Teufel2009, Anetsberger2009,
Borkje2010, Yamamoto2010, Westphal2011}. One approach proposed by
Verlot {\it et al.}\,\cite{Verlot2009} is, instead, to probe the quantum correlation
between the shot noise and the radiation-pressure noise, which, in principle, is totally
immune to thermal fluctuations.

In this article, we analyze a recent experiment performed by Safavi-Naeini
{\it et al.}\,\cite{Safavi-Naeini2012}, in which a radiation-pressure-cooled nanomechanical
oscillator --- the movable mirror of a high-finesse cavity --- is probed by a second beam of light,
detuned from the cavity, for its zero-point mechanical oscillation. The output power
spectrum of the second beam, near the mechanical resonant frequency, serves as an indicator
of the oscillator's zero-point motion. It was experimentally observed that when the second
beam is detuned on opposite sides from the cavity resonance, the output power spectra
turn out to be different. Using theory of linear quantum measurements, we will show that
this experiment not only probes the zero-point fluctuation of the mechanical oscillator at
nearly ground state, but also illustrates vividly the non-trivial correlations between sensing
noise and back-action noise --- an much sought-after effect in the gravitational-wave-detection
community. {\em Its contribution to the output spectrum is equal to the zero-point 
fluctuation for one detuning of the readout beam, and exactly opposite for the other detuning.} 

The outline of this article goes as follows: in Sec.\,\ref{experiment}, we will give a
brief overview of the experiment by Safavi-Naeini {\it et al.}, and present an
analysis of this experiment using quantum measurement theory; in Sec.\,\ref{general},
we will more broadly discuss the nature of mechanical zero-point fluctuation,
show that in attempts to measure the zero-point fluctuation, the contributions from
sensing--back-action noise correlations can generically be comparable to the zero-point
fluctuation itself. In addition, we will discuss linear quantum
measurement devices that use a near-ground-state mechanical oscillator as a probe for
external classical forces near its resonant frequency, and show the limitation on the
measurement sensitivity imposed by the zero-point fluctuation and the connection to 
the SQL; we will conclude in Sec.\,\ref{conclusion}.


\section{A two-beam experiment that measures zero-point mechanical oscillation}
\label{experiment}


In this section, we describe in Sec.\,\ref{subsec:exp} the experiment performed by
Safavi-Naeini {\it et al.}, put its results into the framework of linear quantum
measurement theory in Sec.\,\ref{subsec:frame}, and provide a detailed analysis in
Sec.\,\ref{subsec:detail}. In Sec.\,\ref{subsec:equiv}, we will comment on the
connection between the viewpoint from quantum measurement
and the scattering picture presented in Ref.\,\cite{Safavi-Naeini2012}.

\subsection{Experimental setup and results}
\label{subsec:exp}

\begin{figure}[!b]
\includegraphics[width=0.485\textwidth]{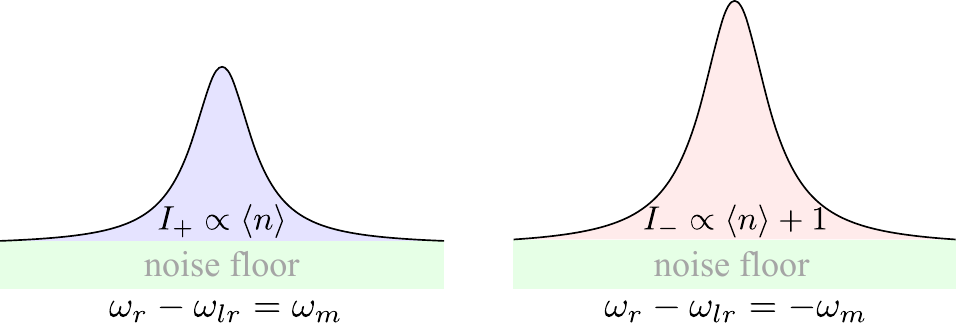}
\caption{Figure illustrating the observed spectra of the readout laser in the
positive-detuning case (left) and the negative-detuning case (right).
\label{fig_asy}}
\end{figure}

In the experiment, two spatial optical modes are coupled to a mechanical vibrational
mode in a patterned silicon nanobeam. One spatial mode --- the cooling mode --- is
pumped with a relatively high power at a ``red'' detuning (lower than resonance),  and
is used to cool the mechanical mode via radiation pressure damping\,\cite{Marquardt2009};
the other cavity mode --- the readout mode --- has a much lower power and
is used for probing the mechanical motion. The readout laser frequency $\omega_{lr}$
is detuned from the resonant frequency $\omega_r$ of the readout mode by either
$+\omega_m$ or $-\omega_m$. The observed spectra of the readout laser are asymmetric
with respect to the detuning $\Delta\equiv\omega_r-\omega_{lr}$. Specifically, in
the positive-detuning case --- $\Delta=\omega_m$, the spectrum has a smaller amplitude
than that in the negative-detuning case. The area $I_{+}$ enclosed by the spectrum in
the positive-detuning case, {\em after subtracting out the noise floor away from the mechanical
resonant frequency}, is proportional to the thermal occupation number $\langle n\rangle$
of the mechanical oscillator, while, in the negative-detuning case, the enclosed area is
$I_-\propto \langle n \rangle+1$. Such asymmetry is illustrated in Fig.\,\ref{fig_asy}.
In Ref.\,\cite{Safavi-Naeini2012}, we introduced the following figure of merit to
quantify the asymmetry:
\be\label{eq_eta}
\eta \equiv \frac{I_-}{I_+}-1=\frac{1}{\langle n \rangle}.
\ee

We interpreted this asymmetry as arising from the quantized motion of the mechanical
oscillator. The asymmetry is thus assigned to the difference between the phonon absorption rate,
proportional $\langle n\rangle$, and the emission rate, proportional to $\langle n\rangle+1$.
This is completely analogous to that used for calibration of motional thermometry of ions/atoms
trapped in electrical/optical traps\,\cite{Diedrich1989, Jessen1992,Monroe1995, Brahms2012}.
Additionally, these scattering processes have an underlying physics similar to bulk nonlinear Raman
scattering processes used in spectroscopic analysis of crystals\,\cite{Hart1970, Hayes_Loudon2004},
where an ensemble of vibrational degrees of freedom internal to the molecular structure of
the system interacts with incident light. Typically in these nonlinear optics experiments, photon
counters are used to keep track of the (anti-)Stokes photons. In contrast, in our experiment, a
heterodyne measurement scheme was used, to find the amplitude quadrature of the readout mode.
Interestingly, by choosing the detuning $\Delta=\pm\omega_m$ and in the resolved-sideband
regime, spectra of the amplitude quadrature are equal to emission spectra of the (anti-)Stokes
photons plus a constant noise floor due to vacuum fluctuation of the light --- the shot noise.
We will elaborate on this point in Sec.\,\ref{subsec:equiv} and show explicitly such a
connection. Intuitively, one can view the cavity mode as an optical filter to selectively
measure the emission spectra --- for $\Delta=\omega_m$, the anti-Stokes
process is significantly enhanced as the emitted photon is on resonance with respect to the
cavity mode, and one therefore measures the spectrum for the anti-Stoke photons; while
for $\Delta=-\omega_m$, the spectrum of the Stokes photon is measured.

\subsection{Interpretation in terms of quantum measurement}
\label{subsec:frame}

Here we  provide an alternative viewpoint to Ref.~\cite{Safavi-Naeini2012},
emphasizing on the role of quantum back-action and its relation to quantization
of the mechanical oscillator. First of all, we separate the experimental system into
two parts. The cooling mode, the mechanical oscillator, and the environmental thermal
bath the oscillator couples to (the left and middle boxes in Fig.~\ref{fig_chart}) together is
the first part, which can be viewed as providing an {\it effective mechanical oscillator}
nearly at the ground state, but with a quality factor significantly lower than the intrinsic
quality factor of the mechanical mode.  It is the zero-point fluctuation of this effective
oscillator that we shall be probing.  The second part of the system consists of the
readout mode (the box on the right of Fig.~\ref{fig_chart}), which couples to the
effective oscillator (the first part of the system) through displacement $\hat x$ alone.
The second part provides us with an output $\hat y$, which contains information about
the zero-point fluctuation of the effective mechanical oscillator.

\subsubsection{The Mechanical Oscillator Near Ground State}

\begin{figure}[!t]
\includegraphics[width=0.48\textwidth]{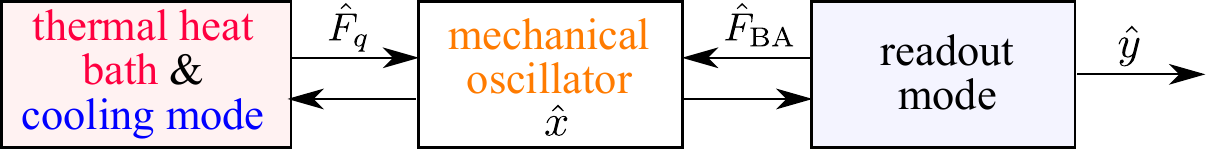}
\caption{Figure illustrating the relation among different parts of the optomechanical
system in the experiment. The thermal heat bath and the cooling mode together
create an effective quantum heat bath for the mechanical oscillator which in turn
couples to the readout mode.
\label{fig_chart}}
\end{figure}

Let us focus on the first part of the system (left two boxes of Fig.~\ref{fig_chart}),
the effective mechanical oscillator (Because this will be a stand-alone subject of study
in later discussions, we shall often ignore the word ``effective''). The environmental
heat bath and the cooling mode together form a {\it quantum} heat bath with fluctuation close
to the zero-point value. In steady state, the ``free" mechanical displacement is
determined by its coupling to this bath (``free" means in absence of the readout mode):
\begin{equation}
\label{zeroreal}
\hat x_q(t) = \int_{-\infty}^t \chi(t-t') \hat F_q(t') {\rm d}t'\,.
\end{equation}
Here $\chi$ is the response function of the mechanical oscillator:
\begin{equation}
\label{chi_time}
\chi(t-t') = -\frac{\left[\hat x(t),\hat x(t')\right]}{i\hbar }
=e^{-\kappa_m |t-t'|/2}\frac{\sin\omega_m (t-t')}{m\omega_m}\,.
\end{equation}
Note that we have an additional decay factor compared with Eq.~\eqref{comm}
which describes an {\it idealized} free oscillator. The decay rate $\kappa_m$ here
is determined jointly by the intrinsic decay rate of the mechanical mode, and the
optomechanical interaction between the mechanical mode and the cooling
mode. The force $\hat F_{q}$ lumps together the fluctuating force acting
on the mechanical mode by the environmental heat bath and the cooling mode.
If the oscillator approaches the ground state only after applying the cooling mode,
then one can show that $\hat F_{q}$ is dominated by fluctuation of the cooling mode.

The above two equations state that for a realistic mechanical oscillator with
non-zero decay rate, its zero-point fluctuation in the steady state can be viewed as
driven by the quantum heat bath surrounding it. We will returning to this prominent
feature of linear quantum systems later in Sec.\,\ref{subsec:zpf}.

\subsubsection{The Quantum-Measurement Process}

Let us now move on to the second part of the system (right box of Fig.~\ref{fig_chart}),
in which the readout mode serves as a linear position meter that measures the
mechanical displacement. We can rewrite the disturbance
$\hat x_{\rm BA}$ in Eq.\,\eqref{eqy2} in terms of the back-action force
$\hat F_{\rm BA}$ arising from radiation-pressure fluctuation
of the readout mode, namely,
\be\label{eq_xt}
\hat x_{\rm BA}(t)=  \int_{-\infty}^t \chi(t-t') \hat F_{\rm BA}(t'){\rm d}t'\,.
\ee
We have assumed that the readout mode does not modify the dynamics of the
oscillator, which is a good approximation for the low pumping power used in
the experiment. Written in the frequency domain, the readout mode
output $\hat y$ [cf. Eq.\eqref{eqy2}] is
\be\label{eq_y}
\hat y(\omega)=\hat z(\omega) +\chi(\omega)\hat F_{\rm BA}(\omega)+
\chi(\omega)\hat F_{q}(\omega).
\ee
where
\begin{equation}
\chi (\omega) =-\frac{1}{m(\omega^2-\omega_m^2+i\kappa_m \omega)}
\end{equation}
is the Fourier transform of $\Theta(t)\chi(t)$, with
$\Theta$ the Heaviside function, i.e., the positive half of $\chi(t)$ (even though
$\chi(t)$ exists for both $t>0$ and $t<0$). The spectral density $S_{yy}(\omega)$
of $\hat y$ then reads:
\be \label{eq_Syy0}
S_{yy} = S_{zz}+2{\rm Re}[\chi^*S_{zF}]+|\chi|^2 S^{\rm BA}_{FF}
+|\chi|^2 S_{FF}^{q}.
\ee
Here these single-sided spectral densities are defined in a symmetrized way (see Appendix~\ref{appa}),
which guarantees bilinearity for the cross spectrum and positivity for self spectrum;
$S_{zz}$ and $S^{\rm BA}_{FF}$ are the sensing-noise and back-action force noise
spectrum, respectively; $S_{zF}$ is the cross correlation between $\hat z$ and $\hat F_{\rm BA}$;
the force spectrum of the effective quantum heat bath made up by the environmental heat
bath and  the cooling mode is given by:
\be
S_{FF}^{q}=(4\langle n\rangle +2)\hbar m \kappa_m \omega_m\,,
\ee
and $\langle n\rangle$ is the thermal occupation number.

\subsubsection{Asymmetry between Spectra}
\label{subsubsec:qualitative}

Experimentally, it was observed that the output spectra $S_{yy}$ for the two
opposite detunings, $\Delta  =\pm \omega_m$, are different --- given the same
thermal occupation number for the
oscillator,
\be
S_{yy}(\omega)|_{\Delta=-\omega_m}\neq S_{yy}(\omega)|_{\Delta=\omega_m}.
\ee
As we will show in the Sec.\,\ref{subsec:detail} that follows, when we flip the
sign of the detuning $\Delta$ of the readout beam, the only term in $S_{yy}$
that changes is $S_{zF}$ --- the correlation between the sensing noise and the
back-action noise. According to Eq.~\eqref{eq_Szx}, we have
\be\label{eq_SzF}
S_{zF}(\omega)\approx -i\,\hbar \frac{\omega}{\Delta},
\ee
in the resolved-sideband regime with the cavity bandwidth $\kappa_r\ll \omega_m$,
which is the case in the experiment. The asymmetry
factor defined in Eq.\eqref{eq_eta} is given by:
\be
\eta = \frac{\displaystyle 2\int  \mathrm{Re}[\chi^* (S_{zF}^{-}-S_{zF}^+)]
{\rm d}\omega}{\displaystyle \int\left[|\chi|^2 S_{FF}^q +  2\mathrm{Re}
 (\chi^* S^+_{zF}) \right]{\rm d}\omega }=\frac{1}{\langle n\rangle}.
\ee
Here $S_{zF}^{\pm}$ is defined by $S_{zF}^{\pm}\equiv S_{zF}|_{\Delta=\pm \omega_m}$,
and in particular around the mechanical resonant frequency $\omega_m$,
where $S_{zF}^{\pm}$ contribute to the above integral,
\be\label{eq_SzF2}
S_{zF}^{\pm}\approx \mp i\,\hbar.
\ee

The asymmetry, or effect of quantum correlation $S_{zF}$, is most prominent when
the thermal occupation number approaches zero. Indeed, if we focus on the quantum
fluctuation by taking the limit of $\langle n \rangle \rightarrow 0$,
we obtain
\begin{equation}
\left.\int  2\mathrm{Re} (\chi^* S^{\pm}_{zF})  {\rm d}\omega\right.=
\mp \int |\chi|^2S_{FF}^{q}|_{\langle n\rangle=0}\,{\rm d}\omega\,.
\end{equation}
In other words, at the quantum ground state, contribution of the quantum correlation
$S_{zF}$ to the readout spectrum $S_{yy}$ has the same magnitude as that of the
zero-point fluctuation, while the sign of the correlation term depends on the sign of
the detuning of the readout beam. {\it This means not only has the experiment probed the
zero-point fluctuation of the mechanical oscillator, it has also demonstrated non-trivial
correlations between sensing noise and back-action noise at the quantum level}.

\subsection{Detailed theoretical analysis}
\label{subsec:detail}

In this section, we supply a detailed calculation of the quantum dynamics and the output
spectrum of the experiment. The dynamics for a typical linear optomechanical device
has been studied extensively in the literature\,\cite{Marquardt2007, Wilson-Rae2007,
Genes2008a}; however, they have been focusing on quantum state of the mechanical
oscillator in ground-state cooling experiments, instead of treating the optomechanical
device as a measurement device. Here we will follow Ref.\,\cite{Miao2010b} and
derive the corresponding input-output relation --- the analysis is the same as the one of
quantum noise in a detuned signal-recycling laser interferometer which can be mapped
into a detuned cavity\,\cite{Chen1,Chen2,scaling_law}. We will focus only on the
interaction between the readout cavity mode and the mechanical oscillator --- the cooling
mode and the thermal heat bath is taken into account by the effective dynamics of the
oscillator as mentioned earlier.

The Hamiltonian of our optomechanical system can be written as\,\cite{Marquardt2007,
Wilson-Rae2007, Genes2008a}:
\begin{align}\label{eq_Hami}
\hat {\cal H}&=\hbar\,\omega_r \hat a^{\dag}\hat a+\hat {\cal H}_{\kappa_r}
+\hbar\,G_0 \hat x\hat a^{\dag}\hat a+\frac{\hat p^2}{2m}+\frac{1}{2}m\,\omega_m^2\hat x^2
+ \hat {\cal H}_{\kappa_m}\,.
\end{align}
Here the first two terms describe the cavity mode including its coupling to the external
continuum; the third term is the coupling between the cavity mode and the mechanical
oscillator; $G_0=\omega_r/L_c$ is the coupling constant with $L_c$ the cavity length;
the rest of the terms describes the dynamics of the effective oscillator (left and middle
boxes in Fig.~\ref{fig_chart}), with $\hat {\cal H}_{\kappa_m}$ summarizing the
dynamics of the cooling mode and the thermal heat bath, as well as their coupling with
the original mechanical oscillator.

In the rotating frame at the laser frequency, the {\em linearized} equations of motion for
the perturbed part --- variation around the steady-state amplitude --- read:
\begin{align}
m(\ddot{\hat x}+\kappa_m \dot {\hat x}+\omega_m^2 \hat x) &=\hat F_{\rm BA}+\hat F_{q},\\
\dot{\hat a}+(\kappa_r/2+i\Delta)\hat a &=-i\,\bar G_0 \hat x+\sqrt{\kappa_r}\,\hat a_{\rm in},
\label{eom2}
\end{align}
where the back-action force $\hat F_{\rm BA}$ is defined as:
\be
\hat F_{\rm BA}\equiv-\hbar\,\bar G_0(\hat a+\hat a^{\dag}),
\ee
and we introduce $\bar G_0=\bar a G_0$ with $\bar a$ being the steady-state amplitude
of the cavity mode and $\hat a_{\rm in}$ is the annihilation operator of the input vacuum
field. The cavity output $\hat a_{\rm out}$ is related to the cavity mode by:
\be
\label{inout}
\hat a_{\rm out}=-\hat a_{\rm in}+\sqrt{\kappa_r}\, \hat a\,.
\ee
with $\kappa_r$ the decay rate (the bandwidth) of the readout mode.

In the steady state, these equations of motion can be solved more easily in the frequency domain.
Starting from the mechanical displacement, we get
\be
\hat x(\omega)= \chi(\omega)[\hat F_{\rm BA}(\omega)+\hat F_{q}(\omega)]\,.
\ee
Here we have ignored modification to the mechanical response function $\chi$ due to the
readout mode---a term proportional to $\bar G_0^2$, assuming that the pumping power
is low. For the cavity mode, we invert Eq.~\eqref{eom2} and obtain
\be
\hat a(\omega)=\frac{{\bar G_0\,\hat x(\omega)+i\sqrt{\kappa_r}\,\hat a_{\rm in}(\omega)}}
{{\omega-\Delta+i\kappa_r/2}}\,,
\ee
which leads to
\be\label{eq_Frp}
\hat F_{\rm BA}(\omega)=\frac{2\,\hbar\,\bar G_0\sqrt{\kappa_r/2}
[(\kappa_r/2-i\omega)\hat v_1+\Delta\, \hat v_2]}{(\omega-\Delta+i\kappa_r/2)
(\omega+\Delta+i\kappa_r/2)}\,,
\ee
with $\hat v_1\equiv (\hat a_{\rm in}+\hat a_{\rm in}^{\dag})/\sqrt{2}$ and
$v_2\equiv (\hat a_{\rm in}-\hat a_{\rm in}^{\dag})/(\sqrt{2}\,i)$ being
the amplitude quadrature and the phase quadrature of the input field, which has
fluctuations at the vacuum level. When combining with Eq.~\eqref{inout}, we
obtain the output amplitude
quadrature
\begin{align}\label{eq_amp}
\hat Y_1(\omega)&=[\hat a_{\rm out}(\omega)+\hat a_{\rm out}^{\dag}(-\omega)]/\sqrt{2} \nonumber\\
&= \frac{(\Delta^2-\kappa_r^2/4-\omega^2) \hat v_1-\kappa_r\Delta\, \hat v_2
+\sqrt{2\kappa_r}\,\bar G_0\Delta\,\hat x}
{(\omega-\Delta+i\kappa_r/2)(\omega+\Delta+i\kappa_r/2)}\,,
\end{align}
whose spectrum is measured experimentally. We put the above formula into
the same format as Eq.\eqref{eq_y} by normalizing $\hat Y_1$
with respect to the mechanical displacement $\hat x$, and introduce $\hat y(\omega)$
and the corresponding sensing noise $\hat z(\omega)$:
\begin{align}\label{eq_amp2}\nonumber
\hat y(\omega)&=\frac{(\Delta^2-\kappa_r^2/4-\omega^2) \hat v_1-\kappa_r\Delta \hat v_2}
{\sqrt{2\kappa_r}\,\bar G_0\Delta}+\hat x(\omega)\\
&\equiv \hat z(\omega) +\chi(\omega)[\hat F_{\rm BA}(\omega)+\hat F_{q}(\omega)]\,.
\end{align}

Taking single-sided symmetrized spectral density of $\hat y$ (see Appendix~\ref{appa}),
we obtain
\be\label{eq_Syy}
S_{yy}(\omega)=S_{zz}+2\mathrm{Re}[\chi^*S_{zF}]+|\chi|^2[S^{\rm BA}_{FF}
+S_{FF}^{q}]\,,
\ee
where
\begin{align}
S_{zz}(\omega)&=\frac{(\Delta^2-\kappa_r^2/4-\omega^2)^2+\kappa_r^2\Delta^2}
{2{\kappa_r}\bar G_0^2\Delta^2}\,,\\
\label{eq_Szx}
S_{zF}(\omega)&=\frac{\hbar(\kappa_r/2-i\,\omega)}{\Delta}\,,\\
S^{\rm BA}_{FF}(\omega)&=\frac{2\hbar^2\bar G_0^2\kappa_r
(\kappa_r^2/4+\omega^2+\Delta^2)}{(\Delta^2-\kappa_r^2/4-\omega^2)^2
+\kappa_r^2\Delta^2}\,.
\end{align}
Here we have used
\begin{equation}
\langle 0|\hat v_j(\omega) \hat v^{\dag}_k(\omega')|0\rangle_{\rm sym}=
\pi\, \delta_{jk}\, \delta(\omega-\omega')\;(j, k=1,2).
\end{equation}

Indeed, only $S_{zF}$ depends on the sign of detuning and contributes to the asymmetry.
In the resolved-sideband case $\kappa_r \ll \omega_m$ and choosing detuning $|\Delta|=\omega_m$,
$S_{zF}$ can be approximated as the one shown in Eq.\,\eqref{eq_SzF}. For a
weak readout beam, we can ignore $S^{\rm BA}_{FF}$ which is proportional to $\bar G_0^2$,
the output spectra around $\omega_m$ for the positive- and negative-detuning cases can be approximated as
\be\label{eq_Syyf}
S_{yy}(\omega)|_{\Delta=\pm \omega_m} \approx \frac{\kappa_r}{2\bar G_0^2} +
\frac{\hbar \kappa_m (2\langle n\rangle +1 \mp 1)}{2m\omega_m[(\omega-\omega_m)^2
+(\kappa_m/2)^2]}\,.
\ee
As we can see, the contribution to output spectra from the quantum correlation
has the same magnitude as the zero-point fluctuation of the mechanical oscillator, with a sign
depending on the detuning. One can then obtain the dependence of the asymmetry factor
 $\eta$ on $\langle n\rangle$ shown in Eq.\,\eqref{eq_eta}.

Interestingly, even if the quantum back-action term $S^{\rm BA}_{FF}$ is much smaller than
$S_{FF}^q$ and has been ignored, given the weak readout mode used in the experiment, the
asymmetry induced by quantum correlation is always visible as long as $\langle n\rangle$ is small.
In addition, any optical loss in the readout mode only contributes a constant noise background --- that is
symmetric with respect to detuning --- to the overall spectrum; therefore, the asymmetry is very robust
against optical loss, and it can be observed without a quantum-limited readout mode which is the case
in the experiment.

\subsection{Connection with the scattering picture}
\label{subsec:equiv}

In the above, we have been emphasizing the viewpoint of position measurement
and interpreting the asymmetry as due to the quantum correlation between the sensing
noise and the back-action noise. Here we would like to show the connection between
this viewpoint and the scattering picture in Ref.\,\cite{Safavi-Naeini2012} that focuses on
the photon-phonon coupling, and in addition, show how spectra of the amplitude quadrature
measured in the experiment are related to emission spectra of the (anti-)Stokes
photons that would have been obtained if we instead take a photon-counting
measurement.

To illustrate these, we introduce the annihilation operator $\hat b$ for the
phonon through the standard definition:
\be
\hat x \equiv \sqrt{\hbar/(2m\omega_m)} (\hat b + \hat b^{\dag})\,.
\ee
and it satisfies the commutator relation: $[\hat b,\,\hat b^{\dag}]=1$. In the rotating frame
at the laser frequency, the Hamiltonian in Eq.\,\eqref{eq_Hami} after linearization is given
by:
\be
\hat {\cal H}=\hbar \Delta \hat a^{\dag}\hat a +\hat {\cal H}_{\kappa_r}+
\hbar \bar g_0(\hat a+\hat a^{\dag})(\hat b+\hat b^{\dag})+\hbar \omega_m \hat b^{\dag}\hat b
+\hat {\cal H}_{\kappa_m}\,,
\ee
where $\bar g_0\equiv \bar G_0 \sqrt{\hbar/(2m\omega_m)}$. The third term is
the photon-phonon coupling: $\hat a^{\dag}\hat b$ describes the anti-Stokes process --- the
absorption of a phonon is accompanied by emission of a higher-frequency photon;
$\hat a^{\dag}\hat b^{\dag}$ describes the Stokes process --- the emission of a phonon is
accompanied by emission of a lower-frequency photon. The photon emission rate of
these two processes can be estimated by using the {\em Fermi's golden rule}. Specifically,
taking into account the finite bandwidth for the photon and phonon due to coupling to the continuum,
the emission rate of the anti-Stokes photon at $\omega_{lr} +\omega $ reads:
\begin{align}\nonumber
\Gamma_{AS}(\omega) &= {\bar g_0}^2 \int {\rm d}\tau\, e^{i\omega\tau}{\cal D}(\omega)\langle
\hat b^{\dag}(\tau)\hat b(0) \rangle \\&
=\frac{{\bar g_0}^2 \kappa_m \langle n \rangle {\cal D}(\omega)}
{(\omega-\omega_m)^2 +(\kappa_m/2)^2}\,;
\label{eq_Gas}
\end{align}
the emission rate of the Stokes photon at $\omega_{lr} -\omega$ reads:
\begin{align}\nonumber
\Gamma_{S}(\omega) &= {\bar g_0}^2 \int {\rm d}\tau\, e^{-i\omega\tau} {\cal D}(-\omega) \langle
\hat b(\tau)\hat b^{\dag}(0) \rangle \\& = \frac{{\bar g_0}^2 \kappa_m (\langle n \rangle+1){\cal D}(-\omega)}
{(\omega-\omega_m)^2 +(\kappa_m/2)^2}\,.
\end{align}
Here the density of state for the photons is determined by the cavity decay rate and detuning:
\be
{\cal D}(\omega) \equiv \frac{\kappa_r/2}{(\omega-\Delta)^2+(\kappa_r/2)^2}\,.
\ee
Were the cavity bandwidth much larger than the mechanical frequency $\omega_m$, the density
of state ${\cal D}(\omega)$ would become flat for frequencies around $\pm \omega_m$, and
we would effectively have a scenario that is similar to the free-space Raman scattering as in
those spectroscopic measurements of crystals\,\cite{Hayes_Loudon2004}. By making a
photon-counting-type measurement of the emitted (anti-)Stokes photons, one could observe an
asymmetric spectrum with two peaks (sidebands) around $\omega_r\pm \omega_m$ of which the
profiles are given by the above emission rates. This is also the case for those emission and
absorption spectroscopic measurements in the ions/atoms trapping experiments\,\cite{Diedrich1989,
Jessen1992,Monroe1995, Brahms2012}.

The situation of our experiment is however different from the usual free-space Raman scattering
spectroscopic measurement by the following two aspects: (i) {\em we are operating in the
resolved-sideband regime} where the cavity bandwidth is much smaller than the mechanical
frequency and the photon density of state is highly asymmetric for positive and negative
sideband frequencies depending on the detuning. This basically dictates that we cannot measure
two sidebands simultaneously, and we have to take two separate spectra by tuning the laser
frequency. In the positive-detuning case $\Delta=\omega_m$, the anti-Stokes sideband is
enhanced while the Stokes sideband is highly suppressed, as the photon density of state is peak
around $\omega=\omega_m$; while in the negative-detuning case $\Delta=-\omega_m$,
the situation for these two sidebands swaps; (ii) {\em we are using heterodyne
detection scheme instead of photon counting}, where the outgoing field is mixed with a large
coherent optical field (reference light) before the photodetector, to measure the output
amplitude quadrature, and the signal is linear proportional to the position of the oscillator, as
we mentioned earlier. Interestingly, there is a direct connection between the spectra of amplitude
quadrature measured in the experiment and the photon emission spectra
that are obtained if making photon-counting measurements.
To show this connection, we use the fact that
\be
[\hat Y_1(\omega),\, \hat Y_1^{\dag}(\omega')] = 0\,
\ee
which is a direct consequence of $[\hat y(t),\,\hat y(t')]=0$ ($\hat y$ is equal to $\hat Y_1$
normalized with respect to the mechanical displacement [cf. Eq.\,\eqref{eq_amp2}]),
and we have
\begin{align}\nonumber
&\langle \hat Y_1(\omega) \hat Y_1^{\dag}(\omega')\rangle_{\rm sym} =
\langle  \hat Y_1^{\dag}(\omega') \hat Y_1(\omega)\rangle \\&=
\frac{1}{2} [\langle \hat a_{\rm out}(-\omega') \hat a^{\dag}_{\rm out}(-\omega) \rangle+
\langle \hat a^{\dag}_{\rm out}(\omega') \hat a_{\rm out}(\omega) \rangle]\,.
\end{align}
Take the positive-detuning case $\Delta =\omega_m$ for instance, $\hat a_{\rm out}(-\omega)$
contains mostly vacuum and negligible sideband signal due to suppression of the Stokes sideband
around $\omega_{lr}-\omega_m$ by the cavity, namely,
$
\langle \hat a_{\rm out}(-\omega') \hat a^{\dag}_{\rm out}(-\omega) \rangle \approx 2\pi\,\delta(\omega-\omega').$
The second term gives the emission spectrum for the output photons shown in Eq.\,\eqref{eq_Gas};
therefore, the single-sided spectral density of the output amplitude quadrature reads:
\be
S_{Y_1Y_1}(\omega) = 1+ 2\Gamma_{\rm AS}(\omega).
\ee
By normalizing the spectrum with respect to the mechanical
displacement,  we have
\be
S_{yy}(\omega)|_{\Delta=\omega_m} = \frac{\kappa_r}{2\bar G_0^2}[1+2 \Gamma_{\rm AS}(\omega)]\,.
\ee
Similarly, by following the same line of thought,
we get
\be
S_{yy}(\omega)|_{\Delta=-\omega_m} = \frac{\kappa_r}{2\bar G_0^2}[1+2 \Gamma_{\rm S}(\omega)]\,.
\ee
The above two equations give identical results to Eq.\,\eqref{eq_Syyf}. Therefore,
the output spectra obtained in our heterodyne detection differ from those in the photon-counting measurement
only by a constant noise floor, which originates from vacuum fluctuation of the amplitude quadrature.
After subtracting this noise floor, we simply recover the emission spectra obtained from taking photon-counting
measurement.


\section{General Linear Measurements of the zero-point fluctuation}
\label{general}


Based on the analysis of the specific experiment of Ref.~\cite{Safavi-Naeini2012}
in the previous section, here we comment on the general features of linear quantum
measurements involving reading out zero-point fluctuation of a mechanical oscillator.
We start from discussing nature of the zero-point mechanical fluctuation in
Sec.\,\ref{subsec:zpf}, proceed to discussion of measurements of it in
Sec.\,\ref{subsec:zpf:meas}, and finally end in Sec.\,\ref{subsec:zpf:SQL} which
discusses its effect on sensitivity for measuring external forces and the connection
to the SQL.

\subsection{The nature of zero-point mechanical fluctuation}
\label{subsec:zpf}

First of all, let us take a closer look at the nature of the zero-point fluctuation of
a realistic harmonic oscillator, which consists of a mechanical mode with eigenfrequency
$\omega_m$ and finite decay rate $\kappa_m$. Suppose we initially decouple
the oscillator from its environmental heat bath and turn on the coupling at $t=0$.
In the Heinserberg picture, the position and momentum of the oscillator at $t>0$ will be
\begin{subequations}
\begin{align}
\label{xH}
\hat x_q(t) &=\hat x_{\rm free}(t)  + \int_0^t  \chi(t-t') \hat{F}_{q}(t'){\rm d}t'\,,\\
\label{pH}
\hat p_q(t) &=\hat p_{\rm free}(t)  +m \int_0^t  \dot\chi(t-t') \hat{F}_{q}(t'){\rm d}t'\,,
\end{align}
\end{subequations}
where
\begin{subequations}
\begin{align}
\hat x_{\rm free}(t)&=e^{-\kappa_m t/2}\left[\hat x(0)\cos\omega_m t +\frac{\hat p(0)}
{m\omega_m}\sin\omega_m t \right]\,, \\
\frac{\hat p_{\rm free}(t)}{m\omega_m}&=e^{-\kappa_m t/2}\left[-  \hat x(0)\sin\omega_m t
+\frac{\hat p(0)}{m\omega_m}\cos\omega_m t \right] \nonumber\\
&-\frac{m\kappa_m}{2} \hat x_{\rm free}(t)\,,
\end{align}
\end{subequations}
are contributions from the free evolution of the initial Schr\"odinger operators
(i.e., undisturbed by the environment),  which decay over time, and get replaced by
contributions from the environmental heat bath [integrals on the right-hand side of
Eqs.\,\eqref{xH} and \eqref{pH}].  Note that for any oscillator with non-zero decay rate,
it is essential to have bath operators entering over time, otherwise the commutation relation
between position and momentum:
\be
[\hat x_q(t),\,\hat p_q(t)]=i\,\hbar
\ee
will not hold at $t>0$ because of
\be
\label{commdecay}
[\hat x_{\rm free}(t),\,\hat p_{\rm free}(t)]=i\,\hbar \,e^{-\kappa_m t}\,.
\ee
This dictates that the heat bath must be such that the additional commutator from terms
containing $\hat F_{q}$ exactly compensate for the decay in Eq.\,\eqref{commdecay},
which leads to the quantum fluctuation-dissipation theorem (see e.g., Ref.\,\cite{Gardiner2004}).

It is interesting to note that this ``replenishing'' of commutators has a classical counterpart,
since commutators are after all proportional to the classical Poisson Bracket.  More
specifically, for a classical oscillator with decay, we can write a similar relation for Poisson
Brackets among the position and momentum of the oscillator, plus environmental degrees
of freedom. The replenishing of the position-momentum Poisson Bracket by environmental
ones, in classical mechanics, can also be viewed as a consequence of the conservation of
phase-space volume, following the Liouville Theorem.  A decaying oscillator's phase-space
volume will shrink, and violate the Liouville Theorem --- unless additional phase-space
volume from the environmental degrees of freedom is introduced.

Nevertheless, the definitive quantum feature in our situation is a fundamental scale in the
volume of phase space, which is equal to $\hbar$. Here we note that if  $\kappa_m \ll \omega_m$,
when reaching the steady state with $\hat x_{\rm free}$ and $\hat q_{\rm free}$ decayed away,
we have
\begin{equation}
\Delta x_q \cdot \Delta p_q  \approx {m\omega_m}\int \frac{d\omega}{2\pi}
S_{xx}^{q}(\omega)\,,
\end{equation}
where  $S_{xx}^{q}\equiv |\chi|^2 S_{FF}^{q}$.
Although $S_{xx}^{q}$ depends on the specific scenario,  they are all constrained by
a Heisenberg-like relation of,
\begin{equation}
S_{xx}^{q} (\omega) \ge 2\hbar\, \mathrm{Im} {\chi(\omega)},
\end{equation}
which is a straightforward consequence of the commutation relation in Eq.\,\eqref{chi_time}.
The equality is achieved at the ground state\,\symbolfootnote[4]{A generalization of this to
thermal states will be the fluctuation-dissipation theorem~\cite{Gardiner2004}.}.
This enforces the same Heisenberg
Uncertainty relation:
\begin{equation}
\Delta x_q \cdot \Delta p_q \ge \hbar/2,
\end{equation}
as an ideal harmonic oscillator whose quantum fluctuations arise ``on its own'', instead of
having to be driven by the surrounding environment. Therefore, in the steady state,
the zero-point fluctuation of the mechanical oscillator can be viewed as being imposed by
the environment due to linearity of the dynamics.

\subsection{Measuring the zero-point fluctuation}
\label{subsec:zpf:meas}

Having clarified the nature of quantum zero-point fluctuations of a mechanical
oscillator in the steady state, let us argue that the effects seen in
Ref.\,\cite{Safavi-Naeini2012} are actually generic when one tries to probe
such fluctuations, namely: the correlation between sensing and back-action
noise can be at the level of the zero-point fluctuation itself.

Let us start our discussion here from Eq.\,\eqref{vanishcomm}, namely,
\be\label{eq_com}
[\hat y(t),\,\hat y(t')]=0\,,
\ee
and the fact that $\hat y$ consists of sensing noise, back-action noise, and
finally the zero-point fluctuation of the mechanical oscillator [cf. Eq.\,\eqref{eqy2}]:
\begin{equation}
\label{inoutgen}
\hat y(t) = \frac{\hat z(t)}{\alpha}  + \alpha \int^t_{-\infty} \chi(t-\tau) \hat F_{\rm BA}(\tau)
{\rm d}\tau +\hat x_{ q}(t)\,.
\end{equation}
Here we have added a factor $\alpha$, which labels the scaling of each term as the
measurement strength which is proportional to the square root of the readout beam
power.  Let us assume that Eq.\,\eqref{eq_com} continues to hold for the same set
of $\hat z$ and $\hat F_{\rm BA}$, for a large set of $\alpha$ and $\chi$: basically
the measuring device works for different mechanical oscillators with different
measuring strength.

Since Eq.~\eqref{eq_com} remains valid for all values of $\alpha$, we extract terms
with different powers of scaling, and obtain
\begin{equation}
\label{commzzff}
\left[\hat z(t),\,\hat z(t')\right]=\left[\hat F_{\rm BA}(t),\,\hat F_{\rm BA}(t')\right]=0\,,
\end{equation}
and
\begin{align}
\label{eqzf}
&\int_{-\infty}^{t'} \chi(t'-\tau)\left[\hat z(t),\, \hat F_{\rm BA}(\tau) \right]{\rm d}\tau \nonumber\\
-&\int_{-\infty}^{t} \chi(t-\tau)\left[\hat z(t'),\, \hat F_{\rm BA}(\tau) \right]{\rm d}\tau  \nonumber\\
+& \left[\hat x_{q}(t),\, \hat x_{q}(t')\right]=0\,,\quad\forall\,t,\,t'\,.
\end{align}
This becomes
\begin{align}
\label{czf}
\int_0^{+\infty}  \chi(\tau) \left[C_{zF}(t-\tau) -C_{zF}(-t-\tau)\right]{\rm d}\tau=
-i\,\hbar \chi(t)\,,
\end{align}
for all values of $t$, where we have defined
\begin{equation}
C_{zF} (t'-t)\equiv \left[\hat z(t),\,\hat F_{\rm BA}(t')\right]\,.
\end{equation}
Here the dependence is only through $t'-t$ because the system assumed to be time-invariant.
We also note that since $\hat z$ is an out-going field, $C_{zF}(t'-t)$ must vanish
when $t'-t>0$, otherwise any generalized force applied on the out-going field $\hat z(t)$ ---
detached from the mechanical oscillator --- can still dynamically influence the mechanical
motion at later times (future) through $\hat F_{\rm BA}(t')$,
which violates the causality\,\cite{BK92, Chen2}. As proved in the App.\,\ref{appb}, in order
for Eq.\,\eqref{czf} to be satisfied for all possible response functions of the oscillator, we must have
\be
C_{zF}(t)=-i\,\hbar \delta_-(t)\,,
\ee
where $\delta_-(t)$ is the Dirac delta function with support only for $t<0$.
In other words,
\be\label{eq_zF}
[\hat z(t), \,\hat F_{\rm BA}(t')]=-i\,\hbar \,\delta_-(t'-t)\,.
\ee

Eq.\,\eqref{inoutgen}, plus the commutation relations in Eqs.\,\eqref{commzzff}
and \eqref{eq_zF}, then provides a general description of linear measuring devices
which do not modify the dynamics of the mechanical oscillator --- simply from the
requirement that the out-going field operators at different times must commute
[cf. Eq.\,\eqref{eq_com}].  In particular, non-vanishing commutator
$\left[\hat x_{q}(t),\,\hat x_{q}(t')\right]$, which underlies the existence of the zero-point
fluctuation, is canceled in a simple way by the non-vanishing commutator between
the sensing noise and the back-action noise [cf. Eq.\,\eqref{eq_zF}].

Now turn to the noise content of the output $\hat y(t)$, i.e., the spectrum,
\be
S_{yy}=\frac{S_{zz}}{\alpha^2}+2{\rm Re}[\chi^*S_{ zF}] +
\alpha^2 S_{FF}^{\rm BA}+S_{xx}^q.
\ee
Let us consider experiments with relatively low measurement strength, so that
the first term $S_{zz}/\alpha^2$ from the sensing noise dominates the output noise.
The next-order terms contain: (i) correlation between the sensing noise and the
back-action noise --- $S_{zF}$; and (ii) the mechanical fluctuation --- $S_{xx}^q$.
If we assume nearly ground state for the mechanical oscillator
\begin{equation}
S_{xx}^{q}(\omega) \approx 2\hbar\, \mathrm{Im} \chi(\omega)\,,
\end{equation}
which, for $\kappa_m \ll \omega_m$, gives
\begin{equation}
\int\frac{{\rm d}\omega}{2\pi} S_{xx}^{q}(\omega) \approx  \frac{\hbar }{2 m\omega_m}\,.
\end{equation}
If $S_{ zF}(\omega)$ does not change noticeably within the mechanical bandwidth, then
\begin{equation}
\int\frac{{\rm d}\omega}{2\pi} 2{\rm Re}[\chi^*(\omega) S_{ zF}(\omega)]  \approx
- \frac{1}{2 m\omega_m} \mathrm{Im}S_{zF}(\omega_m)\,.
\end{equation}
Because of Eq.\,\eqref{eq_zF}, the typical magnitude for $S_{zF}$ is naturally\,\symbolfootnote[5]{In general,
the commutator does not impose any bound on the cross correlation. Here, in a strict sense, is an
order-of-magnitude estimate.}
\begin{equation}
|S_{zF}| \sim \hbar\,.
\end{equation}
Therefore, contributions to the output noise from quantum correlation $S_{zF}$ and mechanical
fluctuation $S_{xx}^q$ can generically become comparable to each other when the mechanical
oscillator is approaching the quantum ground state. The result presented in Ref.\,\cite{Safavi-Naeini2012}
therefore illustrates two typical cases of this generic behavior  [cf. Eq.\,\eqref{eq_SzF2}].

\subsection{Measuring external classical forces in presence of zero-point fluctuation}
\label{subsec:zpf:SQL}

Finally, let us discuss the role of zero-point fluctuation in force measurement, when the
mechanical oscillator is used as a  probe of external classical forces not far away from
the mechanical resonant frequency. The force sensitivity of such a linear measurement
device, in terms of spectral density $S_F$, is obtained by normalizing the displacement sensitivity
$S_{yy}$ with respect to the mechanical response function $\chi$:
$S_F\equiv {S_{yy}}/{|\chi|^2}$. Specifically, from Eq.\,\eqref{eq_Syy0}, we have
\be\label{eq_SF0}
S_{F}(\omega)=
\frac{S_{zz}(\omega)}{|\chi(\omega)|^2}+2{\rm Re}\left[\frac{S_{zF}(\omega)}{\chi(\omega)}\right]
+S^{\rm BA}_{FF}(\omega)+S_{FF}^{q}(\omega)\,.
\ee
Because of the commutation relations in
Eqs.\,\eqref{commzzff} and \eqref{eq_zF}, a Heisenberg Uncertainty Relation exists
among the spectral densities of $\hat z$ and $\hat F_{\rm BA}$, and that is
\be\label{eq_hei}
S_{zz}(\omega)S^{\rm BA}_{FF}(\omega)-S_{zF}(\omega)S_{Fz}(\omega)\ge {\hbar^2}\,.
\ee

When the the sensing noise $\hat z$ and the back-action noise $\hat F_{\rm BA}$
are {\em not correlated} --- $S_{zF}=S_{Fz}=0$, we have
\be
S_{zz}(\omega)S^{\rm BA}_{FF}(\omega)\ge {\hbar^2}\,.
\ee
The above inequality represents a trade-off between sensing
noise $\hat z$ and back-action noise $\hat F_{\rm BA}$. Correspondingly, the
force sensitivity will have a lower bound :
\begin{align}\nonumber
S_{F}(\omega)|_{S_{zF}=0}=&
\frac{S_{zz}(\omega)}{|\chi(\omega)|^2}+S^{\rm BA}_{FF}
(\omega)+S_{FF}^{q}(\omega)\\ \label{eq_SF}&\ge \frac{2\hbar}{|\chi(\omega)|}
+ (4\langle n\rangle+2) \hbar \,m \kappa_m \omega_m\,.
\end{align}
If the mechanical oscillator is in its quantum ground state, namely
$\langle n \rangle=0$, we obtain:
\be\label{eq_SF2}
S_{F}(\omega)\ge \frac{2\hbar}{|\chi(\omega)|}
+2\hbar \,m \kappa_m \omega_m \equiv S_{F}^{\rm Qtot}\,.
\ee
The first term is the usual Standard Quantum Limit (SQL) for force sensitivity with mechanical
probes\,\cite{SQL1, BK92}:
\be
\label{fsql}
S_{F}^{\rm SQL}\equiv \frac{2\hbar}{ |\chi(\omega)|}
=2\hbar m\sqrt{(\omega^2-\omega_m^2)^2+\kappa_m^2\omega^2}\,.
\ee
The second term,
\be
\label{fmq}
S_{F}^{\rm zp}\equiv 2\hbar \,m \kappa_m \omega_m\,,
\ee
{\em arising from the zero-point fluctuation due to mechanical quantization},
also limits the sensitivity.  As we can learn from Eqs.\,\eqref{eq_SF0}, \eqref{eq_hei} and
\eqref{eq_SF2}, the quantum limit, can be surpassed, in principle indefinitely, by building up
quantum correlations between the sensing noise $\hat z$ and the back-action noise
$\hat F_{\rm BA}$ --- in practice the beating factor will be limited by the available optical
power and the level of optical losses.  However, the limit imposed by zero-point fluctuation
cannot be surpassed, and it can only be mitigated by lowering $\kappa_m$, i.e., increasing
the mechanical quality factor.

\begin{figure}[!t]
\includegraphics[width=0.42\textwidth]{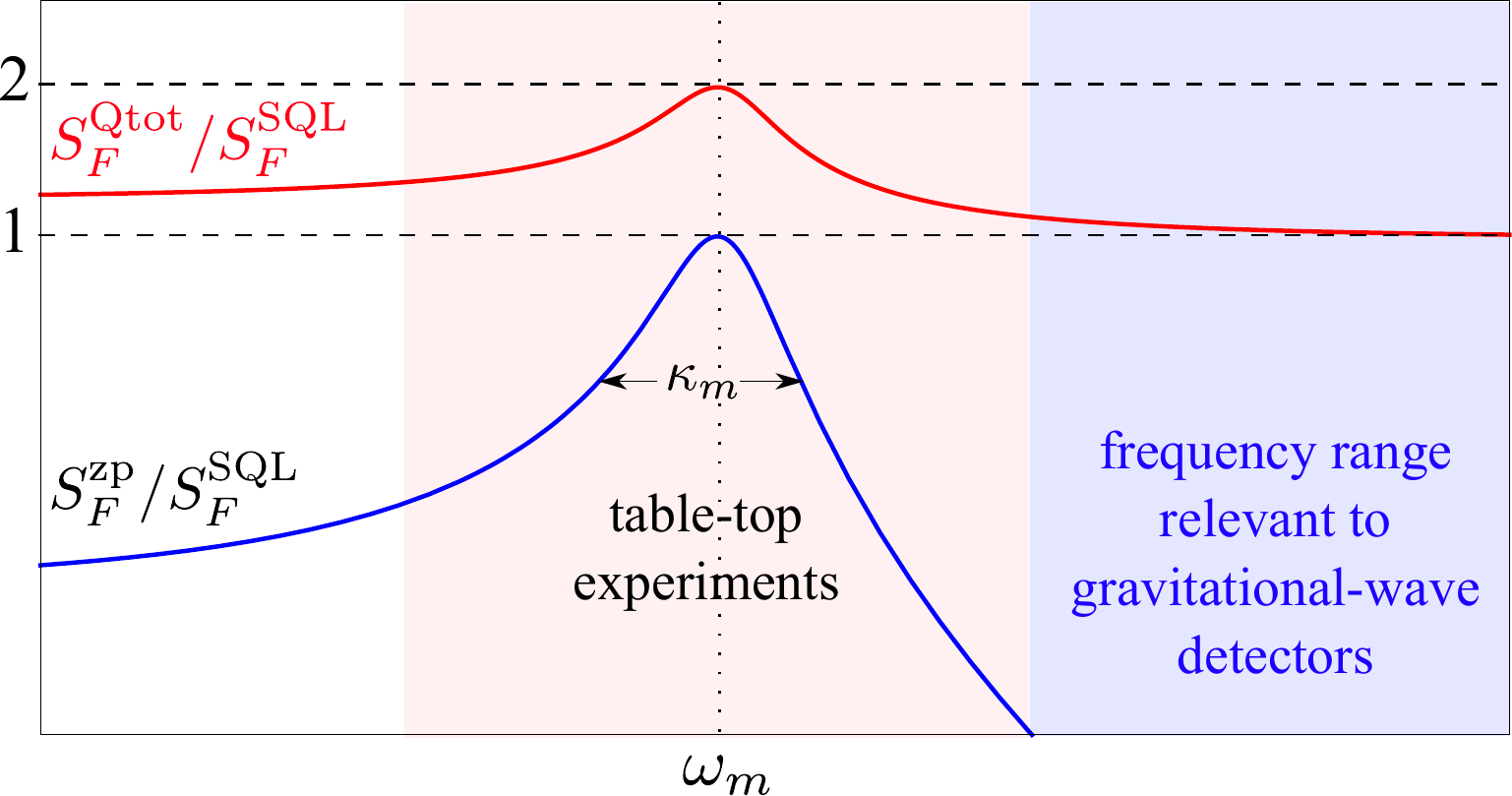}
\caption{(color online) Figure illustrating that total quantum limitation
$S_{F}^{\rm Qtot}$ (red) for force sensitivity and contribution from
zero-point fluctuation $S_{F}^{\rm zp}$ (blue). For clarity, we divide both
by the SQL and use log-log scale.
\label{fig_SQL}}
\end{figure}

\begin{figure}[!t]
\includegraphics[width=0.42\textwidth]{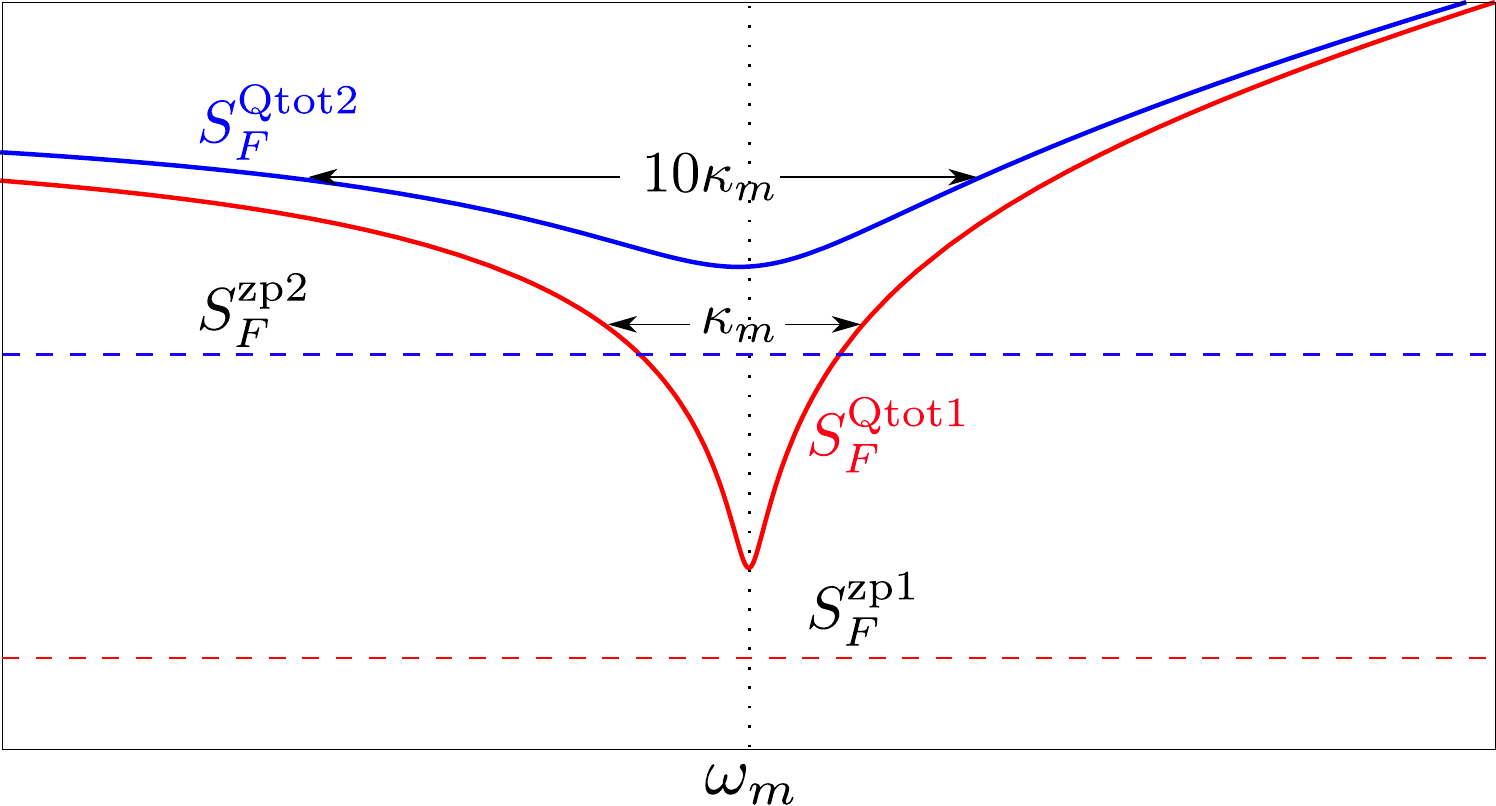}
\caption{(color online) Figure illustrating effect of the mechanical decay rate (bandwidth) $\kappa_m$
on $S_{F}^{\rm Qtot}$ (solid) and $S_{F}^{\rm zp}$ (dashed) --- the larger the mechanical bandwidth,
the lower the force sensitivity (This plot is also in the log-log scale).
\label{fig_force_QL}}
\end{figure}

Braginsky {\it et al.}\,\cite{Braginsky2003} argued that mechanical quantization does not
influence the force sensitivity when measuring classical forces with mechanical probes
--- one only needs to evaluate the quantum noise due to the readout field.  But
these authors had specifically pointed out that they were focusing on ideal mechanical probes with
infinitely narrow bandwidth ($\kappa_m\rightarrow 0$)  and observations outside of that frequency
band. This is close to the actual situation of free-mass gravitational-wave detectors, in which
the mechanical oscillator is the differential mode of four mirror-endowed test masses hung
as pendulum with eigenfrequencies around 1\,Hz and very high quality factor, while the
detection band is above 10\,Hz, well outside the mechanical resonance. Indeed, from Eqs.\,\eqref{fsql}
and \eqref{fmq}, we see that the effect of zero-point fluctuation is only significant not far
away from resonance --- which confirms Braginsky {\it et al.}'s result. More specifically, if
$\kappa_m\ll \omega_m$, we can write, for $|\omega -\omega_m| \ll \omega_m$,
\begin{equation}
S_F^{\rm SQL}
\approx  S_F^{\rm zp}
\sqrt{1+\left(\frac{\omega-\omega_m}{\kappa_m/2}\right)^2}.
\end{equation}
In particular, the limit imposed by zero-point fluctuation is equal to SQL on resonance, and becomes less
important as $|\omega -\omega_m|$ becomes comparable to or larger than the half
bandwidth $\kappa_m/2$, as illustrated in Fig.\,\ref{fig_SQL}. Note that on an absolute scale:
$S_F^{\rm SQL}(\omega)$ is lower near the mechanical resonance,
while $S_F^{\rm zp}(\omega)$ is independent from frequency; at any frequency, lowering
$\kappa_m$, while fixing $\omega_m$ and keeping the oscillator at ground state, always
results in lower noise, as illustrated in Fig.\,\ref{fig_force_QL}. Suppose we are free to
choose from ground-state mechanical oscillators with different $\omega_m$ and $\kappa_m$
as our probe, and that we are always able to reach the SQL at all frequencies, then: (i) if we
know the frequency content of target signals, we can choose probes
that are closely resonant with the target, and (ii) regardless of signal frequency,
probes with lower $\kappa_m$, or equivalently, higher mechanical quality factor, always
provide better force sensitivity.

\section{Conclusion}
\label{conclusion}


We have shown, within the framework of quantum measurement theory, that the
asymmetry in output spectra observed by Safavi-Naeini {\it et al.} can be explained
as due to the quantum correlation between the sensing noise and the quantum
back-action noise; this experiment therefore provides a clear signature of quantum
back-action onto mechanical systems. More broadly, we have shown that having 
quantum-noise correlations showing up at the same level as the zero-point fluctuations 
is a generic feature of measurements that attempt to measure the zero-point fluctuation. 
We have further shown that when an experimentally prepared ground-state mechanical 
oscillator is used as a probe for classical forces near its resonant frequency, its mechanical 
quantization --- through zero-point displacement fluctuation --- does impose an addition 
noise background. This additional noise vanishes only if the oscillator's bandwidth
approaches zero, i.e., when the oscillator becomes ideal. 


\section{Acknowledgements}
\label{acknowledgement}

F.Ya.K.\ is supported by Russian Foundation for Basic Research grant
No. 08-02-00580-a. and NSF grant PHY-0967049.
H. M., H. Y. and Y. C. are supported by NSF grants PHY-0555406,PHY-0956189, PHY-1068881, as
well as the David and Barbara Groce startup fund at Caltech. The
research of A. S.-N. and O. P. has been supported by the DARPA/MTO
ORCHID program through a grant from AFOSR, and the Kavli
Nanoscience Institute at Caltech. A.S-N also gratefully acknowledges 
support from NSERC. We acknowledge funding provided
by the Institute for Quantum Information and Matter, an NSF Physics
Frontiers Center with support of the Gordon and Betty Moore Foundation.


\appendix


\section{Symmetrized cross spectral density}
\label{appa}

In this article,  as in Ref.~\cite{klmtv}, we use the {\em single-sided} symmetrized cross spectral density,
which, given a quantum state $|\psi\rangle$, is defined between a pair of operators $\hat A$ and $\hat B$ as:
\begin{align}
& S_{AB} (\omega) \delta(\omega-\omega')\equiv \frac{1}{\pi}
\langle \psi |\hat A (\omega)  \hat B^\dagger(\omega') |\psi \rangle_{\rm sym}\nonumber\\
&=\frac{1}{2\pi} \langle  \psi |\hat A (\omega)  \hat B^\dagger(\omega')  +
  \hat B^\dagger(\omega') \hat A(\omega) |\psi \rangle\,.
\end{align}
The symmetrization process here allows us to preserves bilinearity of $\tilde S$ on its
entries, i.e.,
\begin{subequations}
\begin{align}
S_{A,c_1 B + c_2 C} &=c_1^* S_{AB} + c_2^* S_{AC}\,, \\
S_{c_1 A + c_2 B,C} &=c_1 S_{AC} + c_2 S_{BC}\,.
\end{align}
\end{subequations}
More importantly, we can show that
\begin{equation}
\label{saa}
S_{AA}>0
\end{equation}
for any field $\hat A$, even if $\left[\hat A(\omega),\,\hat A^\dagger(\omega') \right]\neq 0$.
The positivity~\eqref{saa} allows us to interpret $S_{AA}$ as the fluctuation variance per unit
frequency band --- as in the classical case.

\section{Commutation relation between $\hat z$ and $\hat F$}
\label{appb}

Defining
\begin{equation}
f(t) \equiv C_{zF}(t) + i\hbar \delta_-(t)
\end{equation}
we convert Eq.~\eqref{czf} into
\begin{equation}
\label{eqnewchi}
\int_0^{+\infty}  \chi(\tau) \left[f(t-\tau) -f(-t-\tau)\right]{\rm d}\tau =0\,.
\end{equation}
Assuming analyticity of the Fourier Transform of $f(t)$, it must be written as
\begin{equation}
\tilde f(\omega)  =  \sum_{k} \frac{f_k}{\omega-\omega_k}
\end{equation}
with $\omega_k$ all located on the upper half of the complex plane (not including
the real axis).  Fourier transforming
Eq.~\eqref{eqnewchi} gives us
\begin{equation}
\label{eq:sumpoles}
\tilde\chi_+(\omega) \sum_k
\left[\frac{f_k}{\omega-\omega_k}- \frac{f_k^*}{\omega-\omega_k^*}\right]=0
\,,\quad \omega\in \mathbb{R}\,.
\end{equation}
Because the set $\{\omega_k\}$ is within the upper-half complex plane (excluding
the real axis), the set  $\{\omega_k^*\}$ must be within the lower-half complex plane
(excluding the real axis) --- and the two sets do not intersect.  For this reason,
Eq.\,\eqref{eq:sumpoles} requires $f_k$ to all vanish, and hence
\begin{equation}
C_{zF}(t) = -i \hbar\,\delta_-(t)\,.
\end{equation}


\end{document}